\documentclass[11pt,twoside]{article}
\usepackage[pdftex]{graphicx}
\usepackage{amsmath}
\usepackage{amssymb}
\usepackage{cite}
\usepackage{setspace}
\usepackage{mathtools}
\makeatletter
\renewcommand\@biblabel[1]{#1.}
\makeatother

\let\OLDthebibliography\thebibliography
\renewcommand\thebibliography[1]{
  \OLDthebibliography{#1}
  \setlength{\parskip}{0pt}
  \setlength{\itemsep}{2pt plus 0.3ex}
}

\begin{document}
\newcommand{\pst}{\hspace*{1.5em}}

\newcommand{\rigmark}{\em Journal of Russian Laser Research}
\newcommand{\lemark}{\em Volume 30, Number 5, 2009}

\thispagestyle{plain}

\label{sh}

\begin{center} {\Large \bf
\begin{tabular}{c}
New entropic inequalities for qudit(spin j=9/2) \\[-1mm]
\end{tabular}
 } \end{center}

\bigskip

\bigskip

\begin{center} {\bf
$ V. I. Manko^{1,2*}, T. T. Sabyrgaliyev^{2} $
}\end{center}

\medskip

\begin{center}
{\it
$^1$ P.N. Lebedev Physical Institute\\
Leninskii Prospect 53, Moscow 119991, Russia
\smallskip

$^2$Moscow Institute of Physics and Technology (State University) \\
Dolgoprudnyi, Moscow Region 141700, Russia\\
*Corresponding author e-mail: manko@sci.lebedev.ru
}
\smallskip

\end{center}
\begin{abstract}
We consider information characteristics of single qudit state (spin j=9/2), such as von Neumann entropy, von Neumann mutual information. We review different mathematical properties of these information characteristics: subadditivity and strong subadditivity conditions, Araki-Lieb inequality. The inequalities are entropic inequalities for composite systems (bipartite, tripartite), but they can be written for noncomposite systems. Using the density matrix, describing the noncomposite qudit system state in explicit matrix form we proved new entropic inequalities for single qudit state (spin j=9/2). In addition, we also consider the von Neumann information of a qudit toy model as a function of a real parameter. The obtained inequalities describe the quantum hidden correlations in the single qudit system.
\end{abstract}
\section{Introduction}
Quantum systems have different information charactiristics such as von Neumann quantum entropy \cite{1}, Tsallis \cite{2} and Renyi q-entropies \cite{3}, von Neumann information, etc. The entropies characterize the degree of order in the system. All of them are determined by the density matrix. The density matrix diagonal elements can be interpreted as a probability disribution. Every probability distribution is characterized by the Shannon entropy \cite{4}. For classical system maximum value of Shannon entropy corresponds to maximum disorder in the system. Its quantum generalisation which is called the von Neumann entropy has properties which show relation between systems and subsystems entropies. For any bipartite system there exist subadditivity condition and Araki-Lieb inequality \cite{5}. For tripartite quantum systems strong subadditivity condition was proved \cite{6}. These properties give opportunities to obtain new inequalities for the system density matrix diagonal elements. They are known for composite systems. Moreover, emphasizing virtual subsystems similar inequalities can be proved for noncomposite systems. Generic scheme to get the inequalities is given in \cite{7}\cite{8}. The purpose of our work is investigation of qudit (spin j=9/2) state information characteristics and getting new entropic inequalities. All our results are given in explicit matrix form. We study the entropy-information relations for the single qudit (spin j=9/2) state applying approach developped in the works \cite{7}\cite{8}\cite{9}\cite{10}\cite{11}\cite{12}\cite{13}, where analogous problems for other qudit were discussed. In fact the inequalities which we obtain in the work describe the quantum correlations in noncomposite system. These correlations were called "hidden correlations" because these correlations exist in the systems which have no different subsystems interacting each different withother. Correlations in composite systems are interpreted as some dependence of behaviour of degrees of freedom of one one subsystem on the behaviour of the degrees of freedom of another subsystem. In noncomposite system the hidden correlations are not that obvious but the entropic-information inequalities which we consider in this work clarify the presence of such correlations. In fact we map the noncomposite sustem density matrix  onto density matrix of composite system. This tool we demonstrate on example of concrete qudit, i.e. spin-9/2 system or ten-level atomic system.

Our paper is organized as follows.

In sec.2 we discuss the von Neumann entropy and its properties. Sec.3 is devoted to obtaining the new inequalities for entropies of single qudit (spin $j=9/2$). Next section 4 deals with a toy model. We consider one example of qudit $(j=9/2)$ state with certain distribuion-toy model. More specifically, we investigate the von Neumann information of our toy model as a function of real parameter in different ways. Finally, we give our conclusion in sec.5.
\section{Quantum entropy}
\pst
For any quantum system with density matrix ($ \rho =\rho ^{\dagger}, \rho\geq 0,Tr\rho= 1 $) von Neumann entropy defines as
\begin{equation}
  S(\rho)=-Tr \ \rho \ ln \rho.
\end{equation}
It is quantum generalization of classical Shannon information entropy. 
For any bipartite system there is subadditivity condition of the von Neumann entropy:
\begin{equation}
S(\rho_{12})\leq S(\rho_{1})+S(\rho_{2}).
\end{equation}
where $\rho_{12}, \rho_{1}, \rho_{2}$ are the density matrices of the whole system, first and second subsystems,  respectively. Subadditivity condition has physical interpretation. Since entropy is a measure of chaos, total disorder of the system does not exceed a total disorder of subsystems taken separately. 
Analogous inequality for bipartite system is Araki-Lieb inequality:
\begin{equation}
|S(\rho_{1})-S(\rho_{2})|\leq S(\rho_{12}).
\end{equation}
For any tripartite system there is strong subadditivity condition 
\begin{equation}
  S(\rho_{123})+S(\rho_{2})\leq S(\rho_{12})+S(\rho_{23}).
\end{equation}

These inequalities for composite systems can be applied for noncomposite systems separating two subsystems. Spin $j=(N-1)/2$ has $n$ levels, so it means that its density matrix has sizes $n\times n$. If there are $n$ and $m$ such as $N=nm$, the density matrix of spin can be divided into $n^{2}$ boxes $R_{ij}$ with sizes $m\times m$:
\begin{equation}
  \rho=
  \begin{pmatrix}
    R_{11} & R_{12} & R_{13} & \cdots & R_{1n}\\
    R_{21} & R_{22} & R_{23} & \cdots & R_{2n}\\
    R_{31} & R_{32} & R_{33} & \cdots & R_{3n}\\
    \vdots  & \vdots  & \vdots  & \ddots & \vdots\\
    R_{n1} & R_{n2} & R_{n3} & \cdots & R_{nn}\\
  \end{pmatrix}.
\end{equation}
The density matrices of the subsystems in terms of boxes $R_{ij}$ determines as follows:
\begin{equation}
  \rho_{1}=
  \begin{pmatrix}
    TrR_{11} & TrR_{12} & \cdots & TrR_{1n}\\
    TrR_{21} & TrR_{22} & \cdots & TrR_{2n}\\
    \vdots  & \vdots  & \ddots & \vdots\\
    TrR_{n1} & TrR_{n2} & \cdots & TrR_{nn}\\
  \end{pmatrix},\ \rho_{2}=\sum\limits_{i=1}^n R_{ii}.
\end{equation}
This procedure is described in more detail in \cite{7} and \cite{10}.
It is useful to describe quantum correlations using the von Neumann information:
\begin{equation}
  I=S(\rho_{1})+S(\rho_{2})-S(\rho).
\end{equation}
Due to subadditivity condition, the von Neumann information is nonnegative. It is equal to zero if there is no correlations between subsystems and increases with the growth of correlations between them.

\section{Inequalities for qudit (spin $j=9/2$)}
Using (1), (2), (3) and (4) we get new inequalities for qudit state (spin j = 9/2). It is known to be noncomposite system, but these inequalities can be obtained dropping noncomposite system into auxiliary subsystems.
Qutrit state $(spin j = 9/2)$  has $ 2j+1=10 $ levels. Every level is represented with their state vector from Hilbert space $|j,m\rangle$, where m is projections of the spin on the $Oz$ axis, it is $m=-\frac{9}{2}, -\frac{7}{2}, ..., \frac{9}{2}$. The qudit density matrix $\rho_{m,m^{'}}=\langle j,m| \rho | j,m^{'} \rangle$  has sizes $10\times 10$. Using permutation of basic states we can write explicit form of the density matrix in different ways. One can write the density matrix in the form:
\begin{equation}
  \rho = 
	\begin{pmatrix}
	\rho_{00} & \rho_{01} & \rho_{02} & \hdots & \rho_{09}\\
	\rho_{10} & \rho_{11} & \rho_{12} & \hdots & \rho_{19}\\
	\rho_{20} & \rho_{21} & \rho_{22} & \hdots & \rho_{29}\\
	\vdots & \vdots & \vdots & \ddots & \vdots\\
	\rho_{90} & \rho_{91} & \rho_{92} & \hdots & \rho_{99}\\
	\end{pmatrix}
\end{equation}
where every state projection m of spin $j=9/2$ is relabled using next map to designite the density matrix elements in convenient form :
$-\frac{9}{2} \rightarrow 0; \frac{9}{2} \rightarrow 1; -\frac{7}{2} \rightarrow 2; \frac{7}{2} \rightarrow 3; -\frac{5}{2} \rightarrow 4; \frac{5}{2} \rightarrow 5; -\frac{3}{2} \rightarrow 6; \frac{3}{2} \rightarrow 7; -\frac{1}{2} \rightarrow 8; \frac{1}{2} \rightarrow 9.$
The matrix was divided into 25 boxes $R_{ij}$ $(i,j=\overline{1,...,5})$ with sizes $2\times 2$:
\begin{equation}
  R_{ij}=
  \begin{pmatrix}
    \rho_{2i-2;2j-2} & \rho_{2i-2;2j-1}\\
    \rho_{2i-1;2j-2} & \rho_{2i-1;2j-1}
  \end{pmatrix}.
\end{equation}
The density matrices of subsystems can be calculated using (6):
\begin{equation}
R_{1}=
\begin{pmatrix}
  \rho_{00}+\rho_{11} & \rho_{02}+\rho_{13} & \cdots &  \rho_{08}+\rho_{19} \\
  \rho_{20}+\rho_{31} & \rho_{22}+\rho_{33} & \cdots & \rho_{28}+\rho_{39} \\
  \vdots & \vdots & \ddots &  \vdots \\
  \rho_{80}+\rho_{91} & \rho_{82}+\rho_{93} & \hdots & \rho_{88}+\rho_{99} \\  
\end{pmatrix},
  R_{2}=
  \begin{pmatrix}
    \sum\limits_{i=0}^4 \rho_{2i,2i} & \sum\limits_{i=0}^4 \rho_{2i,2i+1}  \\
    \sum\limits_{i=0}^4 \rho_{2i+1,2i} & \sum\limits_{i=0}^4 \rho_{2i+1,2i+1} \\
  \end{pmatrix}.
\end{equation}
The state with the density matrix $R_{1}$ can be associated like a system with 5 levels. Each level can be represented with a state with certain absolute value of the spins projection $|m|$. The level number n ($n=\overline{1,5}$) has $|m|=\frac{11}{2}-n$. It means that spin may be found in state $|j;m \rangle$  or $|j;-m \rangle$ with probability $P(|j;m \rangle$ or $|j;-m\rangle)=(R_{1})_{nn}$.
The density matrix $R_{2}$ corresponds to 2 level system. Every level is represented with definite sign of projection m. First level with probability $P(m<0)=(R_{2})_{11}$ has negative values of m. The system may be found in second level (the level with positive projection) with probability $(R_{2})_{22}$.
  Subadditivity condition of von Neumann entropy and Araki-Lieb inequality \cite{5} for qudit (spin $j=9/2$) take form:
\begin{equation}
  |S(R_1)-S(R_2)|\leq S(\rho) \leq S(R_1)+S(R_2).
\end{equation}  
  To get strong subadditivity condition it is necessary to get the three subsystems. We will counstruct the state of system with artificial three subsystems that corresponds to the matrix $\rho$ (9) using the approach developed in sec. 1 and \cite{7}, \cite{10}. To get the $12\times12$ matrix the density matrix (9) is expanded via adding 0 elements.
\begin{equation}
  \widetilde\rho_{123} =  
	\begin{pmatrix}
	\rho & O_1\\
	O_2 & O_3\\
	\end{pmatrix},
\end{equation}
where $O_1, O_2, O_3$ matrices with zero elements with sizes $10\times2$, $2\times10$ and $2\times 2$ respectively.
The density matrix (14) is divided into 9 boxes with sizes $4\times 4$ and 16 boxes with sizes $3\times 3$ to obtain matrices $\widetilde\rho_{12}, \widetilde\rho_{23}$ and $\widetilde\rho_{2}$. According to (6): 
\begin{equation}
  \widetilde\rho_{12} =
    \begin{pmatrix}
      \rho_{00}+\rho_{11}+\rho_{22} & \rho_{03}+\rho_{14}+\rho_{25} & \rho_{06}+\rho_{17}+\rho_{28} & \rho_{09} \\
      \rho_{30}+\rho_{41}+\rho_{52} & \rho_{33}+\rho_{44}+\rho_{55} & \rho_{36}+\rho_{47}+\rho_{58} & \rho_{39} \\
      \rho_{60}+\rho_{71}+\rho_{82} & \rho_{63}+\rho_{74}+\rho_{85} & \rho_{66}+\rho_{77}+\rho_{88} & \rho_{69} \\
      \rho_{90} & \rho_{93} & \rho_{96} & \rho_{99} 
    \end{pmatrix},
\end{equation}
\begin{equation}
  \widetilde\rho_{23}=
  \left(
  \begin{array}{cc|cc}
      \rho_{00}+\rho_{44}+\rho_{88} & \rho_{01}+\rho_{45}+\rho_{89} & \rho_{02}+\rho_{46} & \rho_{03}+\rho_{47}\\
      \rho_{10}+\rho_{54}+\rho_{98} & \rho_{11}+\rho_{55}+\rho_{99} & \rho_{12}+\rho_{56} & \rho_{13}+\rho_{57}\\ \hline
      \rho_{20}+\rho_{64} & \rho_{21}+\rho_{65} & \rho_{22}+\rho_{66} & \rho_{23}+\rho_{67}\\
      \rho_{30}+\rho_{74} & \rho_{31}+\rho_{75} & \rho_{32}+\rho_{76} & \rho_{33}+\rho_{77}\\
    \end{array}
  \right).
\end{equation}
Applying this technic (dividing the density matrix $\rho_{23}$ into 4 boxes with sizes $2\times 2$) again the density matrices of second subsystems is calculated:
\begin{equation}
  \widetilde\rho_{2} =
    \begin{pmatrix}
      \rho_{00}+\rho_{11}+\rho_{44}+\rho_{55}+\rho_{88}+\rho_{99} & \rho_{02}+\rho_{13}+\rho_{46}+\rho_{57}\\
      \rho_{20}+\rho_{31}+\rho_{64}+\rho_{75} & \rho_{22}+\rho_{33}+\rho_{66}+\rho_{77}\\
    \end{pmatrix},
\end{equation}
Finally, strong subadditivity condition for system spin $j=9/2$ reads as:
\begin{equation}
  S(\widetilde{\rho}_{123})+S(\widetilde{\rho}_{2})\leq S(\widetilde{\rho}_{12})+S(\widetilde{\rho}_{23}).
\end{equation}
We also calculated the density matrices of first and third subsystems:
\begin{equation}
  \widetilde\rho_{1} = 
    \begin{pmatrix}
      \sum\limits_{i=0}^3 \rho_{i,i} & \sum\limits_{i=0}^3\rho_{i,i+4} & \rho_{08} + \rho_{19} \\
      \sum\limits_{i=0}^3 \rho_{i+4,i} & \sum\limits_{i=4}^7 \rho_{i,i} & \rho_{48} + \rho_{59} \\
      \rho_{80}+\rho_{91} & \rho_{84}+\rho_{95} & \rho_{88} + \rho_{99} \\
    \end{pmatrix},
  \widetilde\rho_{3} = 
    \begin{pmatrix}
      \sum\limits_{i=0}^4 \rho_{2i,2i} & \sum\limits_{i=0}^4 \rho_{2i,2i+1} \\
      \sum\limits_{i=0}^4 \rho_{2i+1,2i} & \sum\limits_{i=0}^4 \rho_{2i+1,2i+1}\\
    \end{pmatrix}.
\end{equation}
Extra subadditivity conditions and Araki-Lieb inequalities read as
\begin{equation}
   |S(\widetilde{\rho}_{1})-S(\widetilde{\rho}_{23})|\leq S(\widetilde{\rho}_{123})\leq S(\widetilde{\rho}_{1})+S(\widetilde{\rho}_{23}),
    |S(\widetilde{\rho}_{12})-S(\widetilde{\rho}_{3})|\leq S(\widetilde{\rho}_{123})\leq S(\widetilde{\rho}_{12})+S(\widetilde{\rho}_{3})
\end{equation}
Finally, we consider physical meaning of matrices. Every density matrix corresponds to the state of system with several levels. Each level can be interpreted as a certain set of projections m. The matrices are described in this table:

\begin{tabular}{|c|c|c|c|c|}
\hline
Matrix & Level 1 & Level 2 & Level 3 & Level 4 \\ \hline
$\widetilde{\rho}_{1}$ & $m=-\frac{9}{2};\frac{9}{2};-\frac{7}{2};\frac{7}{2}$ &  $m=-\frac{5}{2};\frac{5}{2};-\frac{3}{2};\frac{3}{2}$ & $m=-\frac{1}{2};\frac{1}{2}$ &  \\ \hline
$\widetilde{\rho}_{2}$ & $m=-\frac{9}{2};\frac{9}{2};-\frac{5}{2};\frac{5}{2};-\frac{1}{2};\frac{1}{2}$ &  $m=-\frac{7}{2};\frac{7}{2};-\frac{3}{2};\frac{3}{2}$ &  &  \\ \hline
$\widetilde{\rho}_{3}$ & $m=-\frac{9}{2};-\frac{7}{2};-\frac{5}{2};-\frac{3}{2};-\frac{1}{2}$ & $m=\frac{9}{2};\frac{7}{2};\frac{5}{2};\frac{3}{2};\frac{1}{2}$ &  &   \\ \hline
$\widetilde{\rho}_{12}$ & $m=-\frac{9}{2};\frac{9}{2};-\frac{7}{2}$ & $m=\frac{7}{2};-\frac{5}{2};\frac{5}{2}$ & $m=-\frac{3}{2};\frac{3}{2};-\frac{1}{2}$ & $m=\frac{1}{2}$  \\ \hline
$\widetilde{\rho}_{23}$ & $m=-\frac{9}{2};-\frac{5}{2};-\frac{1}{2}$ &  $m=\frac{9}{2};\frac{5}{2};\frac{1}{2}$ & $m=-\frac{7}{2};-\frac{3}{2}$ & $m=\frac{7}{2};\frac{3}{2}$\\ \hline
\end{tabular}\\
In the first column there are the density matrices. A table rows describe corresponding matrix from the first column. There are sets of projections of corresponding level in the next non-empty cells of the row.  
\section{Toy model}
Let us consider some probability distribution. In our toy-model spin $j=9/2$ systems density matrix depends on real parameter $p$ ($0\leq p \leq \frac{1}{45}$) and has next diagonal elements:$\rho_{ii}=p\cdot i$ $ (i=\overline{0,8})$, $p_{9}=1-45p$. It's nonzero nondiagonal elements: $\rho_{10}=\rho_{67}=\rho_{78}=-p$, $\rho_{13}=\rho_{14}=\rho_{47}=\rho_{56}=\rho_{79}=p$,$\rho_{15}=0.5p$,$\rho_{24}=pi$, $\rho_{26}=-1.5pi$, $\rho_{28}=0.1p$, $\rho_{58}=0.2p$ and another nonzero elements those can be calculated using hermitance condition $\rho_ij=\rho*_{ji}$. There 100 elements of the density matrix: 40 nonzero elements and 60 zero elements. Applying formulae (11) and (12) for toy model the subsystems density matrices are obtained:
\begin{equation}
  \rho_1(p) =
  \begin{pmatrix}
    3p & p & 0.5p & p & 0 \\
    p & 7p & pi+2p & -1.5pi & 0.1p \\
    0.5p & -pi+2p & 11p & 0 & 0 \\
    p & 1.5pi & 0 & 15p & p \\
    0 & 0.1p & 0 & p & 1-36p\\
  \end{pmatrix},
  \rho_2(p) =
  \begin{pmatrix}
    25p & -2p \\
    -2p & 1-25p\\
  \end{pmatrix}.
\end{equation}
The von Neumann information for this system as a function of real parameter reads as:
\begin{equation}
  I(p)=S(\rho_{1}(p))+S(\rho_{2}(p))-S(\rho(p)).\\
\end{equation}
In accordance (19) and (16) the density matrices of artificial subsystems corresponds to the supplemented matrix read as:
\begin{equation}
  \widetilde{\rho_1}=
  \begin{pmatrix}
    10p & 0.5p-1.5pi & 0\\
    1.5pi+0.5p & 26p & 0\\
    0 & 0 & 1-36p
  \end{pmatrix},
  \widetilde{\rho_2}=
  \begin{pmatrix}
    15p & -p & 0 & p\\
    -p & 1-37p & p & p\\
    0 & p & 10p & -p\\
    p & p & -p & 12p
  \end{pmatrix}.
\end{equation}
The von Neumann information for this division reads as:
\begin{equation}
  \widetilde{I}(p)=S(\widetilde{\rho}_{1}(p))+S(\widetilde{\rho}_{2}(p))-S(\widetilde{\rho}(p))
\end{equation}
In the figure 1 there are graphs of dependences $I(p)$ and $\widetilde{I}(p)$.
\begin{figure}[h]
  \center{\includegraphics[width=0.7\linewidth]{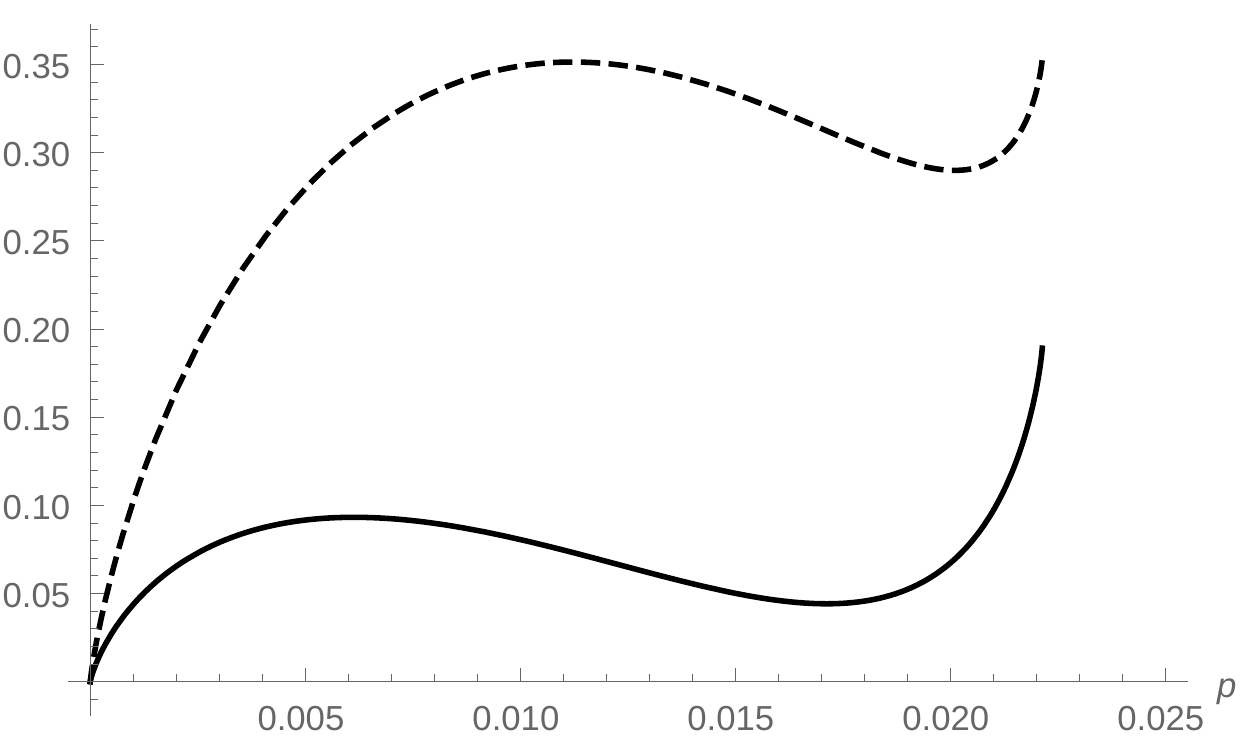}\\}
  \caption{The dependences of the von Neumann informations $I(p)$ (solid line) and $\widetilde{I}(p)$ (dashed line) on real parameter $p$ ($0\leq p \leq \frac{1}{45}$)}
\end{figure}

\section{Summary}
\pst
In conclusion, we point out the main results of our work. We considered single qudit state (spin j=9/2) entropic and information characteristics such as von Neumann quantum information and entropy. Properties of the von Neumann entropy for bipartite and tripartite systems were applied to single qudit (spin j=9/2), although it is known to be the noncomposite system. These properties called subadditivity and strong subadditivity conditions, Araki-Lieb inequality illustrate physical relations between subsystems, so they can be useful for describing hidden quantum correlations in noncomposite systems. To demonstrate this phenomenon we consider toy-model - 10 level system with definate probability distribution. The von Neumann information was investigated as function of probability distribution parameter.The dependence of the information on the parameter determining distribution function reaches maximum and minimum value corresponding different degrees of correlations.
\begin{spacing}{0}

\end{spacing}

\end{document}